\documentclass[12pt]{article}
\usepackage{amsmath}
\usepackage{graphicx,psfrag,epsf}
\usepackage{enumerate}
\usepackage{natbib}
\usepackage{url} % not crucial - just used below for the URL
\usepackage{algorithm}% http://ctan.org/pkg/algorithms
\usepackage{algpseudocode}% http://ctan.org/pkg/algorithmicx
\usepackage{booktabs} 
\usepackage{threeparttable}
\usepackage{tabularx}

\usepackage{amssymb,amsmath,amsthm}
\usepackage{mathtools}

\newtheorem{theorem}{Theorem}
\newtheorem{proposition}{Proposition}
\newtheorem{corollary}{Corollary}

\newcommand{\blind}{0}

\begin{document}

\def\spacingset#1{\renewcommand{\baselinestretch}%
{#1}\small\normalsize} \spacingset{1}

%%%%%%%%%%%%%%%%%%%%%%%%%%%%%%%%%%%%%%%%%%%%%%%%%%%%%%%%%%%%%%%%%%%%%%%%%%%%%%

\if0\blind
{
	\title{\bf Optimal Covariate Weighting Increases Discoveries in High-throughput Biology}
	\author{Mohamad Hasan\thanks{Mohamad Hasan, email: shakilmohamad7@gmail.com}\hspace{.2cm}\\
		\small Department of Statistics, University of Georgia, Athens, GA, email: mshasan@uga.edu\\
		\small and \\
		Paul Schliekelman \\
		\small Department of Statistics, University of Georgia, Athens, GA, email: pdschlie@uga.edu
	\date{September 2017}}
	\maketitle
} \fi

\if1\blind
{
  \bigskip
  \bigskip
  \bigskip
  \begin{center}
    {\LARGE\bf Dynamic Regularized Bayes Borrowing Leveraging Efficiency of Estimation}
\end{center}
  \medskip
} \fi

\bigskip
\begin{abstract}
The large-scale multiple testing inherent to high throughput biological data necessitates very high 
statistical stringency and thus true effects in data are difficult to detect unless they have high effect 
sizes. One promising approach for reducing the multiple testing burden is to use independent information to 
prioritize the features most likely to be true effects. However, using the independent data effectively is 
challenging and often does not lead to substantial gains in power. Current state-of-the-art methods sort 
features into groups by the independent information and calculate weights for each group. However, when true 
effects are weak and rare (the typical situation for high throughput biological studies), all groups will 
contain many null tests and thus their weights are diluted, and performance suffers. We introduce Covariate 
Rank Weighting (CRW), a method for calculating approximate optimal weights conditioned on the ranking of 
tests by an external covariate. This approach uses the probabilistic relationship between covariate ranking 
and test effect size to calculate individual weights for each test that are more informative than group 
weights and are not diluted by null effects. We show how this relationship can be calculated theoretically 
for normally distributed covariates.  It can be estimated empirically in other cases. We show via simulations 
and applications to data that this method outperforms existing methods by as much as 10-fold in the rare/low 
effect size scenario common to biological data and has at least comparable performance in all scenarios.
\end{abstract}

\noindent%
\vfill

\newpage
\spacingset{1.45} % DON'T change the spacing!
\section{Introduction}
The rise of high throughput techniques has dramatically increased the resolution of biological data, often 
approaching the maximum possible for the data type (e.g., all genes in an organism or all nucleotide 
positions in a genome). The promise of this data is that it allows researchers to test very large numbers of 
biological features for effects of interest cheaply and easily. However, although this opens many new 
opportunities for discoveries, it creates a “needle in haystack” problem. Most features are not relevant to 
any particular question at hand, and thus high throughput data requires searching large numbers of features 
for small numbers of effects.

Increasing the number of data features involves a tradeoff: on the one hand, it increases the opportunities 
to find interesting effects; on the other hand, each additional feature decreases the probability of 
successfully detecting effects already present in the data. For this reason, the loss of per-feature 
statistical power often outweighs the benefits of an increased number of data features. Across many data 
types, most effects of interest are weak and cannot be detected with reasonable sample sizes after a large 
multiple testing correction.

One solution to this problem is using other sources of data to prioritize the most promising hypothesis 
tests. With the prevalence of comprehensive databases in many realms of biology, such data sources abound. 
For example, genetic variants in a genome-wide association study could be prioritized using information such 
as 1) markers at nucleotide positions that are more evolutionarily constrained may be better candidates than 
positions that are less evolutionarily constrained; 2) genetic markers that lie near genes that have 
previously been found to be differentially expressed for a trait of interest may be good candidates for being 
causative for the trait; 3) markers that had suggestive (but not conclusive) evidence for association to a 
trait in past studies might be good candidates in future studies.

In a concrete example, \citet{kim2015prioritizing} were looking for genetic markers linked to body weight in 
mice.  They used the Mouse Gene Expression Database to identify $37$ genes whose expression had been 
associated with body weight in previous studies. A marker that is causative for expression for such a gene is 
a good candidate for being causative for body weight. Thus, they used a gene expression data set to find a 
score for the strength of effect of each marker on the $37$ genes. A high score indicates strong evidence 
that the marker is causative for expression for some of the $37$ bodyweight genes and thus is a promising 
target. This score was used as a covariate to prioritize markers and increase discoveries. Such covariates 
exist across many types of data and are easily identified by researchers in those areas.

However, it remains an open question how to use external data best to prioritize data features. Weighted 
p-values have been one of the most widely used approaches.  In p-value weighting, weights are calculated for 
each p-value, with larger weights for more promising tests. 

Consider a situation in which there are $m$ hypothesis tests, and we want test the null hypothesis 
$H_{0i}:\varepsilon_i=0$ against the alternative hypothesis $H_{ai}:\varepsilon_i>0;i=1,\ldots,m,$ where the 
effect size is $\varepsilon_i=\sqrt{n_i}\mu_i/\sigma_i$ for the $i$th alternative mean $\mu_i$ with 
corresponding sample size $n$ and standard deviation $\sigma$. For illustrative purpose, we consider only 
one-sided test; however, one can easily generalize to two-sided test. Define a set of p-values 
$P=\{p_1,\ldots,p_m\}$. If $R$ and $H_0$ refer to the set of rejected null and true null hypothesis, 
respectively, then $R$ controls the Family Wise Error Rate (FWER) if $P(R \cap H_0 ) \le \alpha$. 

Next, define a a set of non-negative weights $w=\{w_1,\ldots,w_m\}$. Then, the simple Bonferroni weighted 
procedure \citep{rosenthal1983ensemble,genovese2006false} rejects a null hypothesis $i$ if
$$
i \ \epsilon \ R=\Big\{i:\frac{P_i}{w_i} \le \frac{\alpha}{m}\Big\},
$$
This weighting scheme will control FWER if the average weight is $1$ \citep{roeder2009genome}. Similarly, 
\citet{genovese2006false} showed that a weighted FDR procedure can be easily conducted by using weighted 
p-values in the procedure of \citet{benjamini1995controlling}.

The larger the weight, the smaller the weighted p-value and thus the higher chance of rejecting the null 
hypothesis. If the weights are chosen well, then the probability of discovering true effects can be 
increased. Generally, the goal is to have higher weights for more promising tests and lower weights for less 
promising tests. Several authors \citep{roeder2006using, rubin2006method, roeder2009genome} have taken the 
approach of finding the weights that maximize the average power across tests, given the true effect sizes. 
Then, the optimal weight for a test with effect size $\varepsilon_i$ is
\begin{equation}\label{eq:rdw_wgt}
	\hat w_i= \Big(\frac{m}{\alpha}\Big)\bar\Phi \Big(\frac{\varepsilon_i}{2} + \frac{c}{\varepsilon_i}\Big) 
	I(\varepsilon_i>0),
\end{equation}
where c is a constant so that $\sum_{i=1}^{m}\hat w_i=m$ and $\Phi=1-\bar \Phi$ is the standard normal CDF.

Unfortunately, expression (1) is not directly applicable because it requires knowledge of the effect sizes. 
These are not generally known - if they were known, there would be no need to conduct the experiment. 
\citet{rubin2006method} proposed using a data splitting approach to estimate the effect sizes and weights. 
One can randomly split the data into two parts and use the first part as a training set to estimate 
$\varepsilon_i$ and the corresponding optimal weights. These are then applied to the testing set. 
Unfortunately, they showed that the power gain from this weighting procedure cannot compensate for the loss 
of power resulting from the split data.

A solution is to incorporate the independent information discussed above in calculating the weights. Now, 
suppose that each hypothesis test $i$ has an associated covariate value $Y_i$, where $Y_i$ is higher for 
tests that are more likely to be true effects. Various authors have proposed weighting based on such external 
information \citep{satagopan2002two, westfall2004weighted, ionita2007genomewide, roeder2009genome, 
dobriban2015optimal, kim2015prioritizing}. \citet{roeder2009genome} (RDW) proposed a method of estimating 
weights by breaking the hypothesis tests into groups based on the covariate, estimating effect sizes for each 
group from the data, and then using \eqref{eq:rdw_wgt} to calculate weights for each group. To maintain FWER 
control, the group sizes must be large enough so that individual null features with chance large test 
statistics will not inflate estimates of effects, thus boosting themselves erroneously. 

Recently, \citet{ignatiadis2016data} introduced an innovative method that they term independent hypothesis 
weighting (IHW). Hypothesis tests are split into groups by an independent covariate that is believed to 
provide information about statistical properties of the hypothesis tests. Weights are calculated for each 
group based on a computational approach that maximizes the number of rejections while maintaining a 
pre-specified FDR. A key component of IHW is that the hypothesis tests are split into $k$ folds. For each 
fold, the IHW optimization procedure is applied to the other $k-1$ folds, and the resulting weights are 
applied to the remaining fold. Thus, the weights for each fold do not depend on p-values in the fold itself. 
Without this procedure, the FWER control would be lost because a chance small p-value in a group would tend 
to inflate the weight and thus erroneously elevate itself to significance.

Although this method is powerful, we will demonstrate that IHW, RDW, and other group-based methods have 
diminished performance in the case that true effects are uncommon (e.g., under $20\%$ of all tests) or the 
power to detect them is low (i.e., most high throughput datasets). Then, true effects of interest are 
difficult to distinguish from the background of null tests.  When groups are formed according to the 
covariate, even the best-ranked groups will have many null tests. Because weights are based on group 
properties, the weights will be diluted by the effect of the null tests. Unfortunately, very low power is 
almost universal in high throughput biological data. Strong signals are rare and usually already discovered, 
and the high multiple testing correction intrinsic to high throughput data makes power low for detection of 
anything but strong signals.

An alternative approach is to weight groups based using a mathematical function based on some criterion that 
is assumed to provide good weighting. \citet{westfall2004weighted} proposed the use of certain data-driven 
quadratic forms. \citet{ionita2007genomewide} proposed an exponential weighting scheme for tests sorted by an 
external covariate. Each subsequent group is twice the size of the previous one, and each weight is $1/4$ of 
the previous one. \citet{kim2015prioritizing} proposed weights based on a vector $\bar n$ that defines group 
boundaries among covariate-sorted p-values and uses weights $w_j=m/\lambda n_{k(j)}$, where $n_{k(j)}$ is the 
smallest value from the vector $\bar n$  that is greater than $j$, and $\lambda$ is a coefficient that makes 
the weights average to one. These approaches are based on predetermined mathematical functions and are not 
data-driven. The advantage of this is that the weights are not based on groups and do not have the problem of 
diluted weights that the RDW and IHW methods suffer. On the other hand, the distribution of effect sizes in a 
data set may not match well with the distribution of the weights.

In this paper, we present a method for calculating p-value weights, Covariate Rank Weighting (CRW), which is 
not group based but still takes into account properties of the data. CRW is based on covariate information 
and works well in the scenario in which true effects are rare and low. The starting point for CRW is the 
optimal weights of \citet{roeder2006using} and \citet{roeder2009genome}. These oracle weights require 
knowledge of the true effect sizes, which is not a reasonable requirement. In our method, hypothesis tests 
are ranked by a covariate. We derive weights that are approximately optimal conditioned on the covariate 
ranking. Our approach requires knowledge of the probabilistic relationship between true effect size and the 
ranking of tests by an independent covariate. We derive this relationship for the case of normally 
distributed covariates. It can be estimated empirically in other cases.

The information in the weights comes from the relationship between the test effect size and the ranking by 
the covariate. Thus, they are not diluted in high throughput data as group-based weights typically are. We 
will show that CRW weights have as much as a $10$-fold advantage in power in the low power situation common 
to high throughput data and are thus the most powerful available method for integrating covariate information 
into large-scale hypothesis testing.

\section{Methods--Covariate rank weighting}\label{sec:Methods}
\subsection{Power formulation}
We will now describe our approach for p-value weighting. Suppose there are total $m$ hypothesis tests with 
$m_1$ true effects and $m_0 = m-m_1$ null effects. The goal is to test the null hypothesis 
$H_{0i}:\varepsilon_i=0$ against the alternative hypothesis $H_{1i}:\varepsilon_i>0;i=1,\ldots,m$. There is a 
test statistic $Z_i$ associated with each hypothesis test;  we assume that this test statistic follows a 
normal distribution. In addition, each hypothesis test $i$ has an associated covariate $Y_i$. This covariate 
is calculated from some independent data that is believed to contain information about the effect sizes of 
the hypothesis tests in the sense that hypothesis tests with larger covariate values are more likely to be 
true alternative tests. We rank the hypothesis tests by the covariate so that each hypothesis test has a 
covariate rank $r_i$. We assume that we know the distribution $f(\varepsilon \mid r_i)$, the probabilistic 
relationship between the covariate ranking and the effect size. We can calculate this directly under certain 
distributional assumptions, or estimate it from data. Our goal is to derive p-value weights based on this 
quantity. Define $\beta(r_i;w_i)$ as the power for the test $i$ with rank $r_i$ and weight $w_i$:
\begin{equation}\label{eq:intPower}
	\beta(r_i;w_i)=P\big(Z_i > Z_{\frac{\alpha w_i}{m}}\mid r_i\big).
\end{equation}

We assume that a significance will be determined using a weighted Bonferroni threshold $\alpha w_i/m$. Next, 
we incorporate the random effect size $\varepsilon$ into \eqref{eq:intPower} and apply Bayes rule:

\begin{equation*}\label{eq:powerLikelihood 1}
	\beta(r_i;w_i )=\frac{\int{P\big(Z_i > Z_{\frac{\alpha w_i}{m}} \mid r_i,\varepsilon\big)P\big(r_i \mid 
	\varepsilon \big) f\big(\varepsilon\big)I(\varepsilon > 0)d\varepsilon}} {P\big(r_i\big)},
\end{equation*}
where $P(r_i)$ is the prior probability of the covariate rank, $P(r_i \mid \varepsilon )$ is the rank 
probability conditioned on the effect size $\varepsilon$, $f(\varepsilon)$ is the probability density 
function of the effect size, and $I(\varepsilon>0)$ is equal to one when the effect size $\varepsilon>0$ and 
zero otherwise. If the test statistic $Z_i$ is normally distributed, then $ P\big(Z_i > Z_{\frac{\alpha 
w_i}{m}}|r_i,\varepsilon\big) = \bar\Phi\big(Z_{\frac{\alpha w_i}{m}}-\varepsilon\big)$. 
$\bar\Phi\big(\cdot\big)$ is equal to one minus the normal CDF. The prior probability of the covariate rank 
is given by $P(r_i )=1/m$, because, lacking additional information, all tests are equally likely to receive 
any covariate rank. This yields the average power

\begin{equation}\label{eq:powerLikelihood 2}
	\frac{1}{m}\sum_{i=1}^{m}\beta(r_i;w_i )=\frac{1}{m}\sum_{i=1}^{m}\int{\bar\Phi\big(Z_{\frac{\alpha 
	w_i}{m}}-\varepsilon\big)mP(r_i \mid \varepsilon)f(\varepsilon)I(\varepsilon > 0)d\varepsilon}.
\end{equation}
We will derive expressions for $P(r_i \mid \varepsilon )$ later in the paper. 

\subsection{Weight}
In the following theorem we show the proposed approximate optimal weight.
\begin{theorem}\label{proposed_weight}
	Given $P\big(r_i \mid E(\varepsilon)\big)$, the optimal weights that maximize the average power 
	(\ref{eq:powerLikelihood 2}) are
	\begin{equation}\label{eq:ContWeight} 
		w_i(r_i; \delta) \approx \Big(\frac{m}{\alpha}\Big) \bar \Phi \Bigg (\frac{E(\varepsilon)}{2} + 
		\frac{1}{E(\varepsilon)} log\bigg(\frac{\delta}{\alpha P\big(r_i \mid E(\varepsilon \mid 
		\varepsilon>0) \big)}\bigg)\Bigg), 
	\end{equation}
	where $\delta \ge 0$ is the normalizing constant such that $\sum_{i=1}^{m}w_i(r_i;\delta)=m.$
\end{theorem}
$E(\varepsilon \mid \varepsilon>0)$ refers to the expected value of the effect sizes for the true effects. In 
order to preserve control of type I error, the average weight must be equal to $1$, i.e. 
$\sum_{i=1}^{m}w_i=m$. The goal is to find the vector of weights $\bar w$ that maximizes the average power 
subject to this constraint. Consider the weight $w_k$ for a specific test $k$. We need to maximize the 
expression below with respect to $w_k$:
\begin{equation*}\label{eq:Likelihood}
	L(w_i;r_i) =\frac{1}{m}\sum_{i=1}^{m}\int{\bar\Phi\big(Z_{\frac{\alpha w_i}{m}}-\varepsilon\big)mP(r_i 
	\mid \varepsilon)f(\varepsilon)I(\varepsilon > 0)d\varepsilon} - 
	\delta\Big(\frac{1}{m}\sum_{i=1}^{m}w_i-1\Big),
\end{equation*}
where $\delta$ refers to the Lagrange multiplier. We maximize by differentiating with respect to $w_k$, 
setting equal to zero, and solving. After performing simple algebra, this produces
\begin{equation}\label{eq:finalLikelihood}
	\int\left(e^{Z_{\frac{\alpha w_k}{m}\varepsilon}-\frac{\varepsilon^{2}}{2}}\right)P(r_k \mid 
	\varepsilon)f(\varepsilon)d\varepsilon=\frac{\delta}{\alpha}.
\end{equation}
The above integration is not easily tractable, except for special cases of $P(r_i \mid \varepsilon)$ and 
$f(\varepsilon)$. Numerical solutions of the weights are tractable. However, we obtain an approximate general 
weight expression that will be applicable in a wide variety of situations. Let
\begin{equation*}\label{eq:taylorFunction}
	g(\varepsilon)=\left( e^{Z_{\frac{\alpha w_k}{m}\varepsilon}-\frac{\varepsilon^{2}}{2}}\right) P(r_k \mid 
	\varepsilon).
\end{equation*}	

Equation (\ref{eq:finalLikelihood}) can then be interpreted as $E(g(\varepsilon))$. Since $\varepsilon$ is a 
random variable and $g(\varepsilon)$ is a differentiable function, then $E(g(\varepsilon))= 
g(E(\varepsilon))$ in the first order Taylor series expansion of $E(g(\varepsilon))$. Consequently, after 
simple algebraic manipulation, the approximate version of the weight (\ref{eq:ContWeight}) is obtained. With 
this approximation, the weights maximize the average (over tests) power evaluated at the expected non-zero 
effect size. Exact weights would maximize the average expectation of power over effect size. 

This is similar to the \citet{roeder2009genome} weight \eqref{eq:rdw_wgt} but incorporates the relationship 
$P(r_i \mid \varepsilon)$. The main advantage of this weight is that it does not require knowledge of the 
effect sizes. Instead, it only requires the expected value of the effect sizes and the relationship between 
the effect size and the covariate rank.

\citet{dobriban2015optimal} generalized the \citet{roeder2009genome} weights to the scenario that effect 
sizes follow a Gaussian distribution with a known mean and variance. They optimized average power in a manner 
similar to our approach. However, their approach only considers the distribution $f(\varepsilon)$. Our work 
builds on theirs by explicitly including the covariate rank $r_i$. A key advantage of our method is that 
using the rank distribution explicitly accounts for the other tests in determining how to weight a given 
test. That is, it uses more information in specifying the weight. These methods will be compared more fully 
in a future manuscript.

We compared the approximate weights and the exact weights obtained by numerical integration (Figure 18 in 
SI). Both the approximate and the exact approaches provided similar results. To compute the approximate 
weights, we used expression (\ref{eq:ContWeight});  and to compute the exact weights, we applied the 
numerical integration to (\ref{eq:finalLikelihood}). We applied these procedures to the two datasets shown in 
Figure \ref{fig:real_data_examples}.

To obtain the optimal value of $\delta$, we applied Newton-Raphson algorithm (NR) when $\varepsilon > 1$ and 
Grid search algorithm when $\varepsilon \ \epsilon \ [0, 1)$. Although Newton-Raphson is computationally 
faster, it is sensitive to a correct guess of the initial value and is also sensitive to non-convexity. When 
the effect size is low, then the weights are flatter, and the NR method often does not converge. Therefore, 
we used grid search algorithm to obtain $\delta$ for low $\varepsilon$. The full algorithm for finding 
$\delta$ is given in the SI (Section 19). 

\subsection{Rank probability given continuous effect \boldmath $P(r_i \mid \tau_i)$}\label{sec:Probability of 
Rank for Continuous}
The primary source of information in the weights is the probability  $P(r_i \mid \varepsilon_i)$, where $r_i$ 
is the rank by the covariate and  $\varepsilon_i$ is the effect size for the hypothesis test $i$.  This 
quantity determines how the weights vary with covariate rank. The covariate is assumed to have an effect size 
$\tau_i$ that is distinct from the test effect size $\varepsilon_i$. Thus, the covariate rank $r_i$ depends 
on the covariate effect size $\tau_i$ and not on the test effect size $\varepsilon_i$ directly. However, the 
fundamental assumption of our approach is that there is a positive relationship between the two effect sizes 
and hence higher values of the covariate indicate more promising tests.

In the following, we will derive the distribution for covariate rank given covariate effect size, $P(r_i \mid 
\tau_i)$. We will then discuss the relationship between test effect size $\varepsilon_i$ and covariate effect 
size $\tau_i$. 

\begin{theorem}\label{CovRank_thm}
	Suppose there are $m$ hypothesis tests with associated covariate values and $r_i, r_{0i}$ and $r_{1i}$; 
	$i=1, \ldots, m$ are the ranks of a random covariate value $t$ given the covariate effect size $\tau_i$ 
	among covariate values for all tests, among the true nulls, and among the true alternatives, 
	respectively. Then the rank probability $P(r_i \mid \tau_i)$ is the expectation of the pmf of the sum of 
	two Binomial random variables $r_{0i}$ and $r_{1i}$, and the expectation is over $t$. If $t$ is from 
	null, then $r_{1i} \sim binomial(m_1,F_1)$ and $r_{0i} \sim binomial(m_0-1,F_0)$, where $F_0$ and $F_1$ 
	are the CDFs of the covariates under the null and alternative hypotheses, respectively. If $t$ is from 
	alternative hypothesis, then $r_{1i} \sim binomial(m_1-1,F_1)$ and $r_{0i} \sim binomial(m_0,F_0)$. The 
	rank probability is given by
\end{theorem}

\begin{equation}\label{eq:ranksProb}
	P(r_i=k \mid \tau_i )=E_T\Big[\sum_{k_0=1}^{k}\left\lbrace P(r_{1i}=k-k_0+1 \mid \tau_i,t )P(r_{0i}=k_0 
	\mid \tau_i,t )\right\rbrace \Big]
\end{equation}

See Appendix 1 for the proof and further discussion. Equation \eqref{eq:ranksProb} is not easily tractable, 
and finding a closed form solution is difficult. However, \eqref{eq:ranksProb} can be solved numerically and 
can also be easily simulated. We solved \eqref{eq:ranksProb} using the importance sampling method of the 
Monte Carlo (MC) simulation. Importance sampling is an MC simulation approach in which the integral is 
expressed as an expectation of a function of a random variable. Then, the density of the random variable is 
chosen appropriately to reduce the variance of the estimated integral.

Furthermore, since the term inside the expectation is the sum of two independent binomial random variables, 
we propose a normal approximation of the term. This approximation introduces a procedure that is faster and 
easier than many other algorithms:
\begin{proposition}\label{CovRankProp}
	Rank probability $P(r_i=k \mid \tau_i )$ is the expected value of the normal PDF, i.e.,
	\begin{equation*}
		P(r_i=k \mid \tau_i) =
		\begin{cases}
			E_T N(k-1 \mid \mu_0,\sigma_0^2), & \text{if } \tau_i=0\\
			E_T N(k-1 \mid \mu_1,\sigma_1^2), & \text{if } \tau_i > 0
		\end{cases},
	\end{equation*}
	where
	\begin{equation*}
		\begin{cases}
			\mu_0 = (m_0-1)(1-F_0 ) + m_1 (1-F_1 ) + 1\\
			\mu_1 = m_0(1-F_0 ) + (m_1-1)(1-F_1 ) + 1\\
			\sigma_0 = (m_0-1)(1-F_0 )F_0 + m_1 (1-F_1 )F_1\\
			\sigma_1 = m_0(1-F_0 )F_0 + (m_1-1)(1-F_1 )F_1
		\end{cases}.
	\end{equation*}
\end{proposition}

$N(k-1 \mid \mu_i,\sigma_i^2)$ represents the normal pdf evaluated at the value $k-1$.  

\subsection{Relationship between test effect and covariate effect}\label{sec:test-cov rel}
For each test $i$, take $\varepsilon_i$  as the effect size for the test statistic and $\tau_i$ as the effect 
size for the covariate. The information in the weights comes from the relationship between the covariate rank 
$r_i$ and the test effect size $\varepsilon_i$, while the covariate rank depends on the covariate effect size 
$\tau_i$. The fundamental assumption of our approach is that there is a positive relationship between the two 
effect sizes. 

In the implementation in this paper, we assume that there is a noisy linear relationship between the test 
effect size and the covariate effect size. That is, 

\begin{equation}\label{eq:CovTestEffect}
	\tau_i=\alpha+\beta\varepsilon_i+N \big(0,\sigma^2 \big)
\end{equation}

for some $\alpha$, $\beta$, and $\sigma$. We conduct a linear regression of estimated covariate effect sizes 
on estimated test effect sizes and then use this to find the corresponding value of the covariate effect (see 
work flow in SI). This value is then used to calculate the rank probability \eqref{eq:ranksProb} used in the 
weight formula \eqref{eq:ContWeight}. The linear model assumption is not fundamental to our approach. It 
would be straightforward to fit the covariate effect - test effect relationship using a non-linear regression 
method. 

Note that the weight formula (\ref{eq:ContWeight}) only depends on $P\big(r_i \mid E(\varepsilon \mid 
\varepsilon>0) \big)$. Thus, we only need to identify the covariate effect size corresponding to the mean 
test effect size among true alternative effects. This is the point at which the regression line is most 
accurate and also not highly sensitive to deviations from a linear relationship. 

In the simulation section, we explore the effect of the noise in the relationship between $\varepsilon_i$ and 
$\tau_i$. In the Supplementary Material, we show simulation results for the case that the relationship 
between test effect and covariate effect is non-linear when we assume that it is linear. A moderate degree of 
non-linearity does not have a large impact on the performance of our method.

\subsection{Weight and rank probability given binary effect $P(r_i \mid \varepsilon)$}
We also propose weights for binary effects, i.e., estimating weights for $H_{0i}:\varepsilon_i=0$ vs. 
$H_{1i}:\varepsilon_i=\varepsilon$ (a fixed value). Therefore, the prior probability of having a true effect 
is $p=m_1/m$.  Following similar arguments to the continuous effect size case, we find the following weights:
\begin{corollary}\label{proposed_weight_bin}
	Given $P\big(r_i \mid \varepsilon)$, the optimal weights that maximize the average power 
	(\ref{eq:powerLikelihood 2}) are
	\begin{equation}\label{eq:BinWeight}
		w_i(r_i;\delta) = \Big(\frac{m}{\alpha}\Big) \bar \Phi \Bigg (\frac{\varepsilon}{2} + 
		\frac{1}{\varepsilon} log\bigg(\frac{\delta m}{\alpha m_1 P(r_i \mid 
		\varepsilon)}\bigg)\Bigg)I(\varepsilon_i = \varepsilon).
	\end{equation}
	where $\delta \ge 0$ is the normalizing constant such that $\sum_{i=1}^{m}w_i(r_i;\delta)=m.$
\end{corollary}
The main advantage of the binary case is that it is fairly easy to calculate, and an exact version of the 
weight is available. One can compute $P(r_i \mid \varepsilon)$ in a manner similar to the continuous case.

\subsection{Work flow}
The full work flow for the CRW method is given in the SI. 

\subsection{Simulation of dilution effect in group-based methods }
We used simulations to explore the performance of group based methods like IHW when true effects are weak and 
rare. We assumed that there were a total of $m=40,0000$ hypothesis tests. The number of true null tests was 
$m_0=m \pi_0$ where $\pi_0$, the proportion of null hypothesis, varied between simulations. The number of 
true alternative tests was $m_1=m-m_0$. The effect sizes for the true alternative tests were generated from 
$N(\mu_\epsilon,1)$, where $\mu_{\epsilon}$ varied for different scenarios. For each hypothesis test, the 
test statistic and the covariate value were generated from independent normal distribution with mean equal to 
the effect size and a standard deviation of $1$. Thus, the test statistic and the covariate were correlated 
across tests due to the shared mean. The tests were then sorted by the covariate and divided evenly into $10$ 
groups. The proportion of true alternative hypothesis tests and the mean effect size were calculated for the 
top group and plotted. Next, we conducted a simulation using the IHW method with $10$ groups. Data were 
simulated in the same manner as above. We used the \textit{ihw} function from the \textbf{R} package 
\textbf{IHW} to calculate weights for each group. The IHW weights were plotted as a function of $\pi_0$ and 
$\mu_{\epsilon}$. $100$ simulation replicates were conducted for each combination of $\pi_0$ and 
$\mu_{\epsilon}$ and the mean weight for each group was calculated across replicates. 

\section{Results}
\subsection{Power in high throughout studies}\label{sec: power high throughput}
\begin{figure}
	% The arguments in the next line are {height}{optional width}{used only by OUP for typesetting}[filename, 
	%in directory art]
	%\figurebox{20pc}{25pc}{}[hightroughputPower.eps]
	\includegraphics[width=5in,height=3.5in]{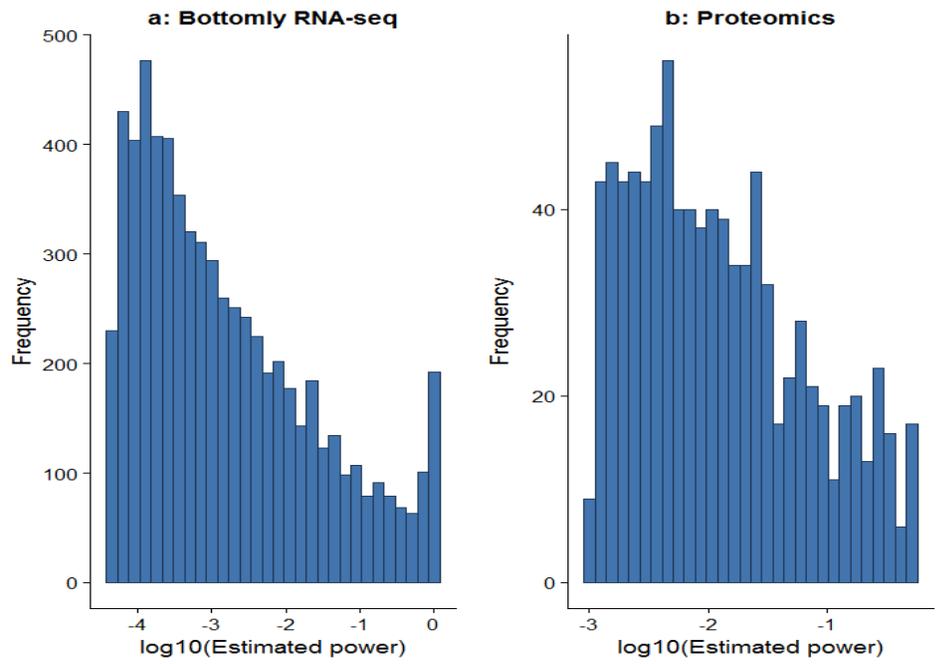}
	%\includegraphics[width=8cm,height=8cm][hightroughputPower.png]
	% note that files may not be rotated
	\caption{Distribution of Statistical Power in high throughput studies. Post-hoc power estimated across 
	top $m_1$ most significant tests in a) the RNA-seq data of \citet{bottomly2011evaluating} and b) the 
	proteomics dataset of \citet{dephoure2012hyperplexing}.}
	\label{fig:hightroughputPower}
\end{figure}

First, we demonstrate that low statistical power is pervasive in high throughput biological data. 
Figure~\ref{fig:hightroughputPower} shows the distribution of estimated power across tests for two different 
datasets: 1) the RNA-seq data of \citet{bottomly2011evaluating} and 2) the proteomics dataset of 
\citet{dephoure2012hyperplexing}. In each dataset, we estimated $m_1$, the number of true effect hypothesis 
tests, using the method of \citet{storey2003statistical}. We then estimated the effect size for the $m_1$ 
most significant tests from their p-values and calculated the corresponding power assuming that the test 
statistic follows a normal distribution and using a Bonferroni correction (see SI for the details). 
Simulations included in the SI (Figure 3.1-3.2) demonstrate that this method is informative about the true 
distribution of power but will tend to overestimate power. Fig.~\ref{fig:hightroughputPower} shows the 
estimated distribution of the power for the two datasets. Even with the likely overestimate of power, the 
estimated values are very low for nearly all tests in both datasets. 

For the RNA-seq data, the median of estimated power values is $0.08\%$ and the $90\%$ quantile is $10\%$. For 
the proteomics data, the median is $1\%$ power and the $90\%$ quantile is $17\%$ power. It should be 
emphasized that these values are only for the tests estimated to be true positive features. This very low 
power is a result of the multiple testing correction. There are many effects in both datasets that would be 
strongly significant if they were tested individually but are undetectable when tested with many other 
features. If we used an FDR-based approach rather than the Bonferroni correction, power for some features 
could be increased by $1-2$ orders of magnitude. Even then, power remains very low for most features. This 
demonstrates two key motivating points for this study: 1) Most effects are not detectable unless their power 
can be boosted in some way; 2) Methods for boosting power using covariates must perform well when there are 
large numbers of low-power tests.

\subsection{Dilution of group gased weights}
\begin{figure}
	\includegraphics[scale=0.7]{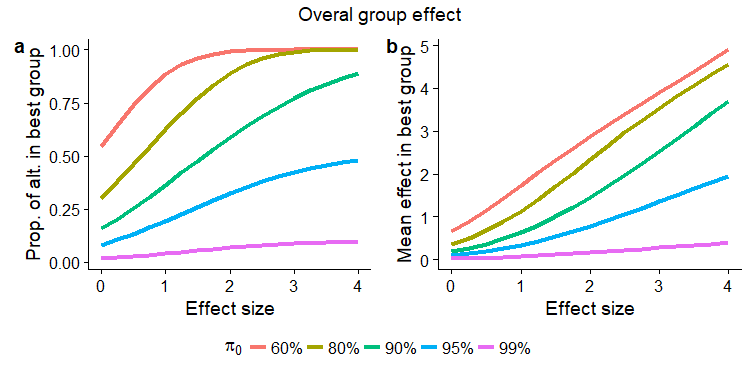}
	\caption{\textbf{Composition of top group}. Simulated test statistics ordered by simulated covariate. The 
	effect sizes for alternative hypothesis tests were generated from a normal distribution with the mean 
	shown on the x-axis. 
		The test statistic and covariate value for each test were generated from a normal distribution with 
		mean equal to the effect size. The left-hand plot shows the proportion of true alternative hypothesis 
		tests in the top-ranked group. The right-hand plot shows the average effect size in the top-ranked 
		group. }
	\label{fig:GroupEffect1}
\end{figure}
\begin{figure}
	\includegraphics[scale=0.6]{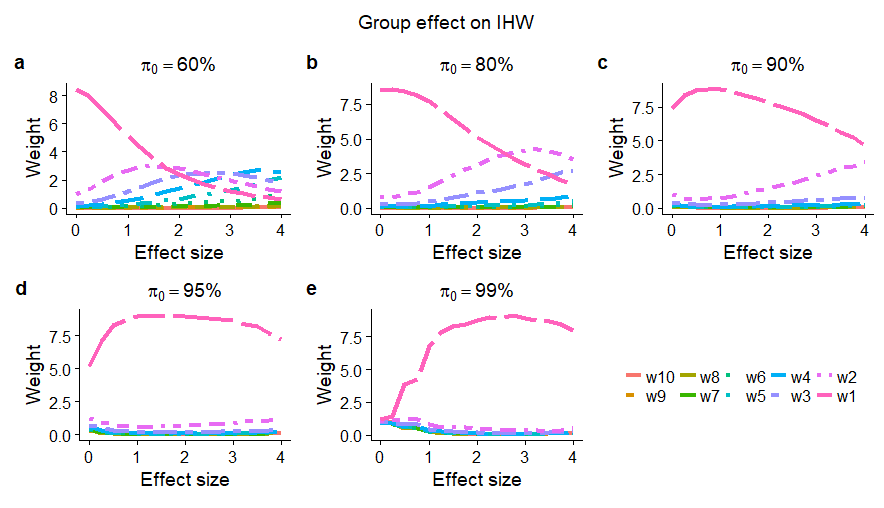}
	\caption{\textbf{IHW weights}. Each curve shows the average weight for one group across 100 replicates. 
	In the legend, $w_1-w_{10}$ refers to the weights from the first to the tenth  group.}
	\label{fig:GroupEffect2}
\end{figure}
We will demonstrate the dilution effect for group based methods discussed in the Introduction. These methods 
separate tests into groups by the covariate and then calculate group weights based on group properties. The 
problem that we now demonstrate is that even the top groups will contain many null tests when true effects 
are weak and rare. As a result, the group weights will be diluted and are less informative.

As described in the Methods, test statistics and covariate values were generated for $40,000$ hypothesis 
tests. Tests were sorted into $10$ groups by the covariate, and the proportion of true alternative tests and 
the average effect size was calculated for the top group (Figure \ref{fig:GroupEffect1}).

Ideally, the top group would be comprised primarily of true effects. However, this is only true when the 
effect sizes are very large, and the proportion of true alternative tests is also large. Over parameter 
ranges that are typical of most high throughput datasets (for example, the proportion of null hypothesis 
$\pi_0 \ge 90\%$ and mean effect size less than one), less than $25\%$ of tests in the top group are true 
alternative hypothesis. The right-hand panel shows the corresponding average effect size in the top group. 
For reasonable values of $\pi_0$, this is much less than the true average effect size of the data.

Figure \ref{fig:GroupEffect2} shows the weights for each group when the IHW method is applied to data 
generated in the same manner (see Methods for details). The IHW method is demonstrated because it is the most 
powerful weighting method currently available. When true alternative effects are weak and rare, only the 
strongest effects have a chance of being significant. In this case, IHW gives the top group the highest 
weight and other groups receive weights that are near one or below. The plot shows that the highest weight 
that the top group receives across all combinations of mean effect size and $\pi_0$ is about eight. As 
$\pi_0$ approaches $100\%$ and the average effect decreases below one, the weight of the top group decreases 
to even lower values. These weights are not large enough to boost power sufficiently for weak effects in high 
throughput data.  We will show in the next section that the CRW method produces stronger weighting for these 
weak effects. 

\subsection{Properties of covariate rank weights (CRW)}
\begin{figure}
	% The arguments in the next line are {height}{optional width}{used only by OUP for typesetting}[filename, 
	%in directory art]
	%\figurebox{20pc}{25pc}{}[ranksProbCont.eps]
	\includegraphics[width=5.7in,height=4.5in]{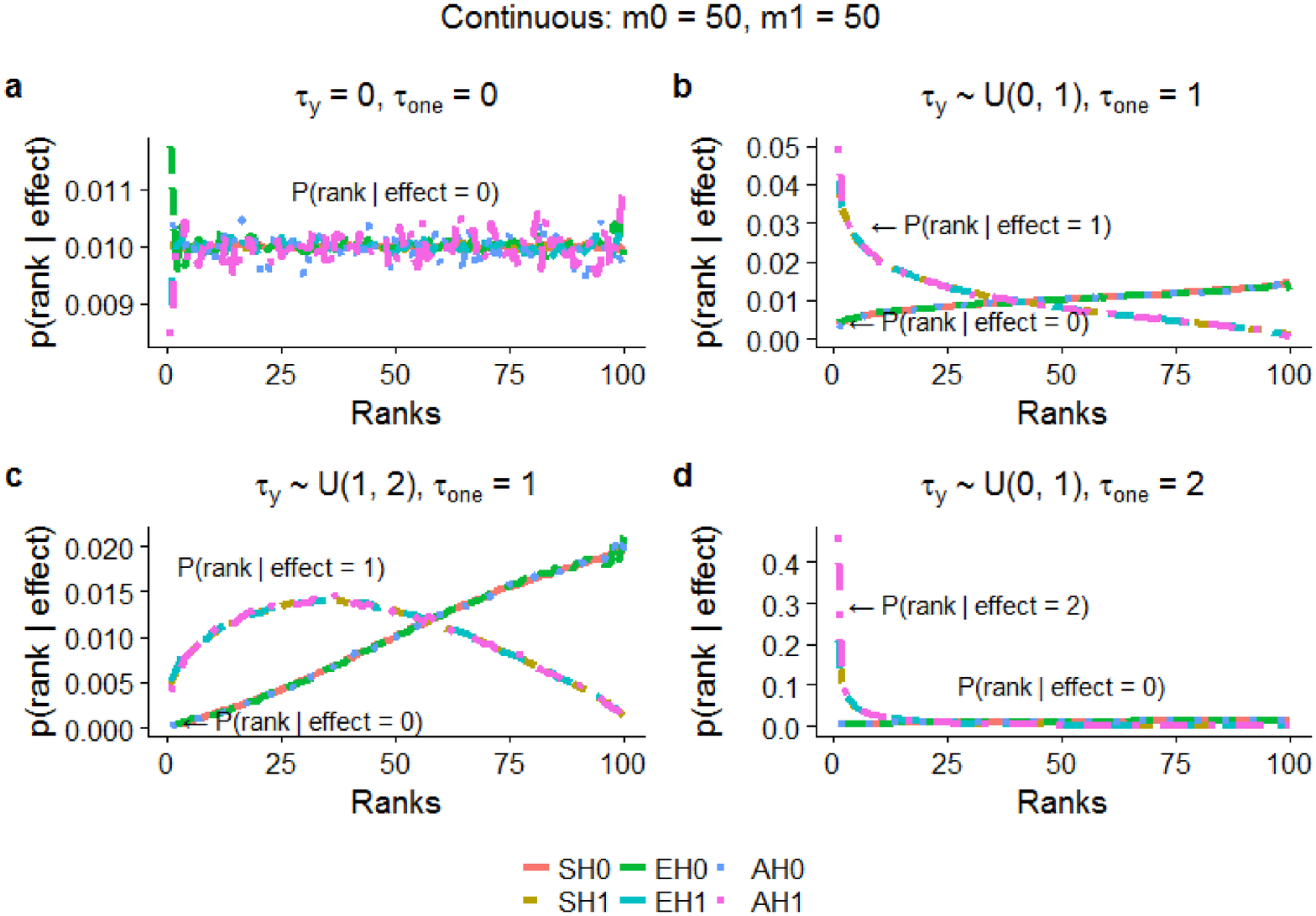}
	% note that files may not be rotated
	\caption{$P(r_i=k \mid \tau)$ for continuous effects for three different scenarios: 1) simulation, 2) 
	exact CRW method, and 3) normal approximation of CRW method. We assumed that there are $m=100$ tests of 
	which $m_0=50$ are true nulls and $m_1=50$ are true effects. The true null tests have mean $0$ and the 
	true effect tests have the distribution $\tau_y$ as shown at the top of each plot. Each plot consists of 
	six curves, SH0, SH1, EH0, EH1, AH0, and AH1, where the first letter represents the method (S = 
	simulated, E = exact, and A = approximate), and H0 and H1 represent the hypothesis type. Three curves 
	(SH0, EH0, and AH0) starting from the bottom-left represent $P(r_i=k \mid \tau=0)$, and the remaining 
	three curves (SH1, EH1, and AH1) starting from top-left represent $P(r_i=k \mid \tau=\tau_{one})$, where 
	$\tau_{one} = \{0,1,2\}$. All the curves show the rank probabilities of a test statistic with effect size 
	either $\tau_i = 0$ or $\tau_i = \tau_{one}$ among all tests. (a) The rank probabilities of a null test 
	when all tests are true nulls. (b) Rank probabilities of a true effect test with effect size 
	$\tau_{one}=1$, when $50$ test statistics are true nulls with effect size $0$, $49$ test statistics have 
	effect size $\tau_y \sim uniform(0,1)$, and one test statistic has effect size $\tau_{one}=1$. Similarly, 
	(c) and (d) show the probabilities for different effect size distribution.}
	\label{fig:ranksProb_cont}
\end{figure}

We explore properties of the rank probabilities and their effect on the CRW weights. 
Figure~\ref{fig:ranksProb_cont} shows the rank probability $P(r_i \mid \tau)$ of a test from three different 
approaches: 1) Simulation approach, 2) Exact numerical solution, and 3) Normal approximation of the proposed 
method. All plots of the simulation suggest nearly perfect alignment with the proposed CRW (exact and 
approximate) methods. This suggests that the normal approximation (\ref{CovRankProp}) is valid.

The values of the weights for different features depend primarily on the value of the rank probability 
$P(r_i=k \mid \tau_i )$. When the rank for a feature has higher probability, then the feature will be 
weighted more highly.  This figure shows how the probability varies for different combinations of effect 
sizes for the target test and the remaining tests. 

If all effects are null, then ranks have a uniform distribution (Figure \ref{fig:ranksProb_cont}a). In 
Fig.~\ref{fig:ranksProb_cont}b, there are $50$ covariates with effect sizes sampled from the $uniform(0,1)$ 
distribution and $50$ with an effect size of zero. A covariate with effect size at the upper end of this 
distribution (effect size $=1$) has its rank probability maximized at rank one, but still has a substantial 
probability of a wide range of lower ranks. This reflects the random variation in covariate statistic values 
about their expected value. Similarly, a covariate with effect size zero has its rank maximized at rank $100$ 
but may appear at any rank. In  Fig.~\ref{fig:ranksProb_cont}c, the distribution for the alternative 
hypothesis effects is shifted upwards to $uniform(1,2)$. The distribution for rank probabilities for effect 
size one now shifts downwards, with rank probability maximized at middle values. In 
Fig.~\ref{fig:ranksProb_cont}d, a covariate with high effect size relative to other tests has a high 
probability of being ranked in the top $10$ and low probability of a lower rank. 

\begin{figure}
	% The arguments in the next line are {height}{optional width}{used only by OUP for typesetting}[filename, 
	%in directory art]
	%\figurebox{20pc}{25pc}{}[probandweightvsrank.eps]
	\includegraphics[width=5.7in,height=4.5in]{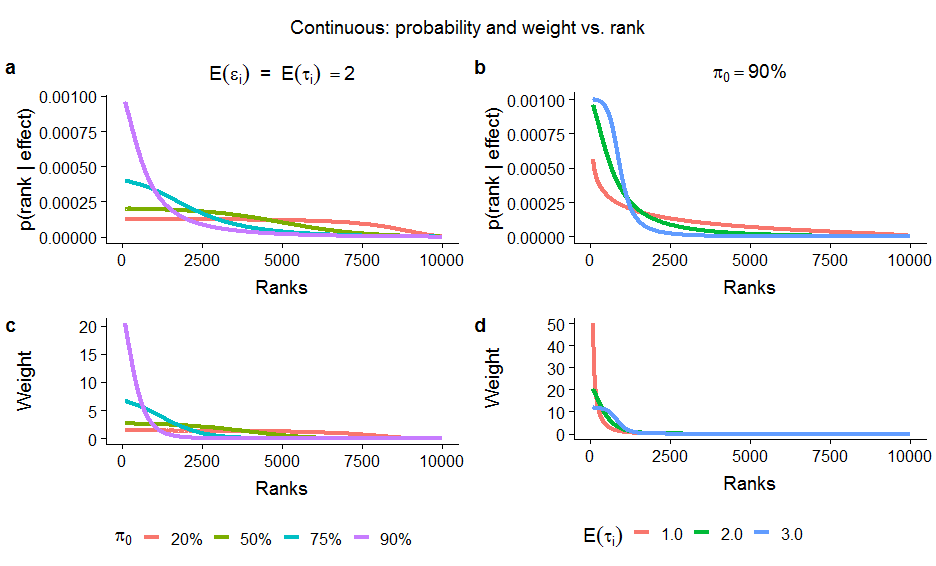}
	% note that files may not be rotated
	\caption{The ranks probabilities, $P(r_i=k \mid E(\varepsilon_i ))$, and the corresponding normalized 
	weights, $w_i$, (a) and (c) for different proportions of the true null hypothesis $(\pi_0)$; and (b) and 
	(d) for different mean covariate-effect $E(\tau_i)$. The effect size was the same for all true effect 
	tests and equal to the value shown. There were $10,000$ total tests. The curves were smoothed to 
	eliminate random variation due to Monte Carlo integration.}
	\label{fig:prob_and_weight_vs_rank}
\end{figure}
Figure~\ref{fig:prob_and_weight_vs_rank} shows ranks probabilities and their corresponding weights. In this 
figure, all hypothesis tests with true effects have the same covariate effect size, and the rank probability 
and corresponding weight is shown for one of these tests. In the left-hand plots, the effect size is fixed at 
a value of $2$, while the proportion of true nulls varies between the different curves. In the right-hand 
plots, the proportion of true nulls is fixed at $90\%$, and the effect size varies between curves. 

As is evident from inspection of the weight equation \eqref{eq:ContWeight}, higher weight is assigned to 
tests that have a higher rank probability. When the true effect tests are more differentiated from the 
background, then there is a higher probability of higher rank and thus higher weight for higher ranked tests. 
This is observed when comparing the different null proportion curves in 
Fig.~\ref{fig:prob_and_weight_vs_rank}a and Fig.~\ref{fig:prob_and_weight_vs_rank}c. When true effect 
hypothesis tests are rare (e.g., $90\%$ null proportion), then a given true effect hypothesis test is highly 
likely to be ranked highly and thus it gets a high weight. When there are many true effect hypothesis tests 
with similar effect sizes (e.g., $20\%$ null proportion), then true effect tests will be spread out over many 
ranks and thus no rank receives high weight. 

Similar trends are seen across varying effect sizes in Fig.~\ref{fig:prob_and_weight_vs_rank}b and 
Fig.~\ref{fig:prob_and_weight_vs_rank}d. When the covariate effect size is high (blue curve), then there is a 
high probability that all true effect tests will be ranked above all true null tests. In this case, all ranks 
in the upper $10\%$ are highly likely to be true effect tests and thus all receive the same intermediate size 
weight. When the covariate effect size is weak (red curve), then null tests are likely to spread over all 
except the very highest ranks. Thus, the highest ranks get high weights because they are highly 
differentiated from the background. 

This behavior is a highly desirable aspect of the CRW method, i.e., true effects that are rare and low in 
size will receive the highest boost from weighting. In contrast, such effects will tend to be washed out in 
group-based approaches to weighting. Comparing Fig.~\ref{fig:prob_and_weight_vs_rank} to 
Fig.~\ref{fig:GroupEffect2}, we see that the highest ranked tests receive much higher CRW weights than IHW 
weights (weights of $50+$ with CRW versus $7-8$ in IHW). 

In practice, the rank probabilities are calculated by assuming that all alternative hypothesis covariates 
have equal effect size. This effect size is set equal to the covariate effect size calculated in section 
\ref{sec:test-cov rel}. That is, the covariate effect size corresponding to the mean alternative hypothesis 
test effect size. This is how the weights are calibrated to the data. The rank probabilities reflect the 
estimated distribution of effect sizes and the weights vary according to the rank probabilities. Plots of the 
rank probabilities are given for many other scenarios in the SI.

\subsection{Power comparisons of competing methods} \label{sec: power comparison}
\begin{figure}
	% The arguments in the next line are {height}{optional width}{used only by OUP for typesetting}[filename, 
	%in directory art]
	%\figurebox{20pc}{25pc}{}[power1.eps]
	\includegraphics[width=5.7in,height=4in]{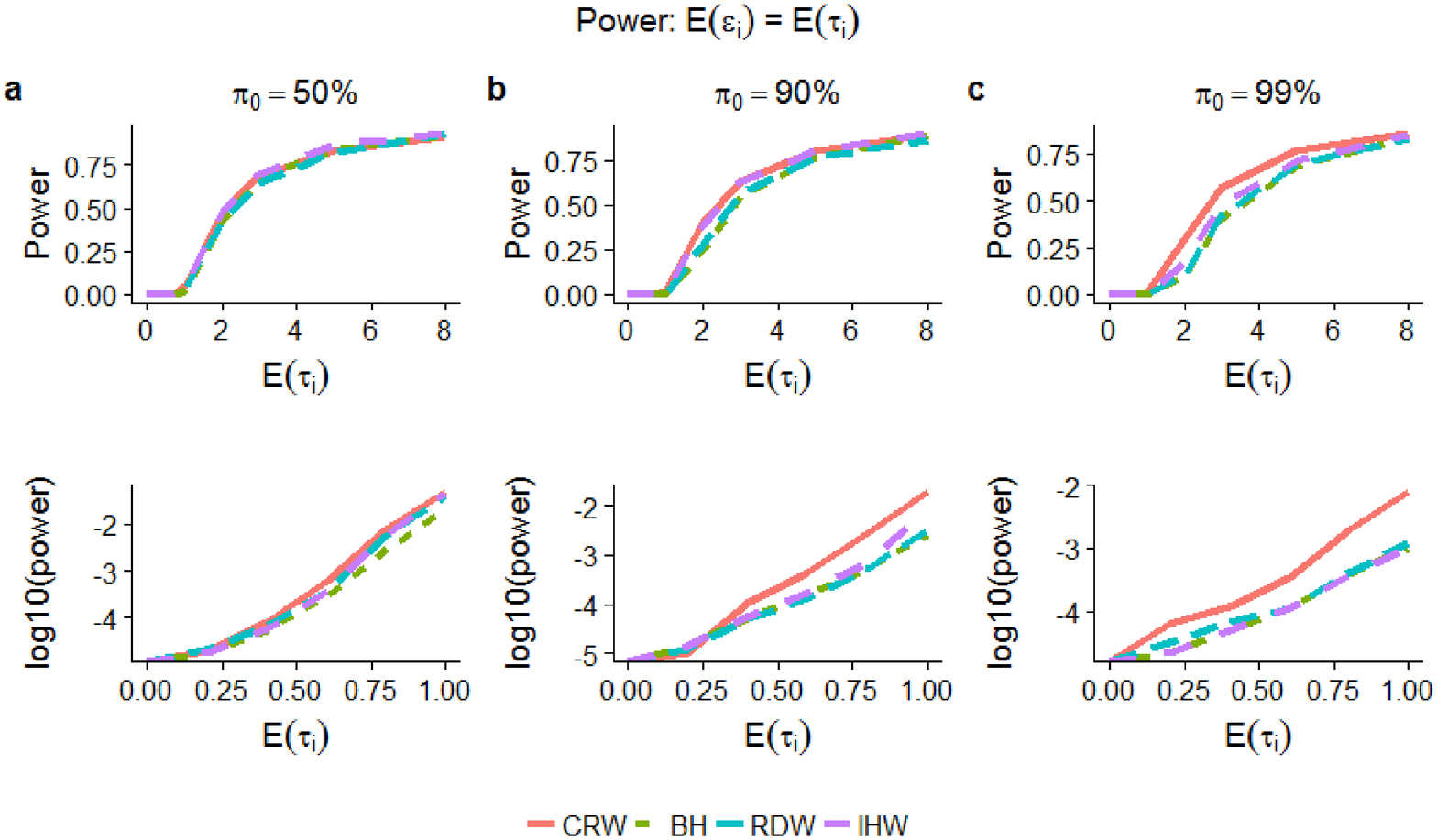}
	% note that files may not be rotated
	\caption{The power of four methods when the mean test-effect $E(\varepsilon_i)$ is equal to mean 
	covariate-effect $E(\tau_i)$. Each plot consists of four curves of CRW, BH, RDW, and IHW methods. The 
	first row shows the power for the full range of effect sizes, and the second row shows $log_{10}(power)$ 
	for low effect sizes. The three columns represent the cases of $50\%,90\%,$ and $99\%$ true null 
	hypothesis.}
	\label{fig:power1}
\end{figure}

Next, we show comparisons of the performance of our method to existing methods of p-value weighting with 
covariate information. We compared the performance of \citet{benjamini1997false} (BH) with no weighting, 
\citet{roeder2009genome} (RDW), and Independent Hypothesis Weighting (IHW) \citep{ignatiadis2016data} to our 
proposed method Covariate Rank Weighing (CRW) in terms of power, FWER, and FDR. To implement the BH and IHW 
methods, we used \textit{p.adjust} and \textit{ihw} function from \textbf{R} software and followed the 
\cite{benjamini1997false} FDR procedure. To implement the RDW method, we applied the weighting procedures 
described in \citet{roeder2009genome}. When the number of the true null hypothesis is low 
(Figure~\ref{fig:power1}a), the IHW method outperforms CRW if the mean effects size is large. However, for 
all other situations, CRW equals or outperforms all of the three methods. 

Unlike IHW, CRW computes the weight for each test; therefore, it can more easily detect true effects even 
though the proportion of the true effects and the effect sizes is relatively low. The relative benefit of CRW 
over other methods is highest when power is low (bottom row of Figure~\ref{fig:power1}). In this case, it can 
have a $10$-fold advantage in power over IHW. As shown in section \ref{sec: power high throughput}, very low 
power is the norm for high-throughput data. Because there are large numbers of tests, the difference between 
power of, e.g. $0.1\%$ and $1\%$ can make a large difference in the number of discoveries. Consider, for 
example, a hypothetical genomewide association study which typically might have $500,000$ tests and $5\%$ 
true effect hypothesis tests. That is, $25,000$ true effect tests. Many of these tests are highly correlated 
and thus suppose that there are on the order of $5,000$ true underlying effects. 
Figure~\ref{fig:hightroughputPower} shows that typical power for these tests might be in the range $10^{-2}$ 
to $10^{-3}$. If we assume $5,000$ tests with an average power of $10^{-3}$, then we would expect on the 
order of $5$ independent significant effects. However, if power can be boosted to $10^{-2}$, then we expect 
$50$ significant effects.

\begin{figure}
	% The arguments in the next line are {height}{optional width}{used only by OUP for typesetting}[filename, 
	%in directory art]
	%\figurebox{20pc}{25pc}{}[power2.eps]
	\includegraphics[width=5.7in,height=4in]{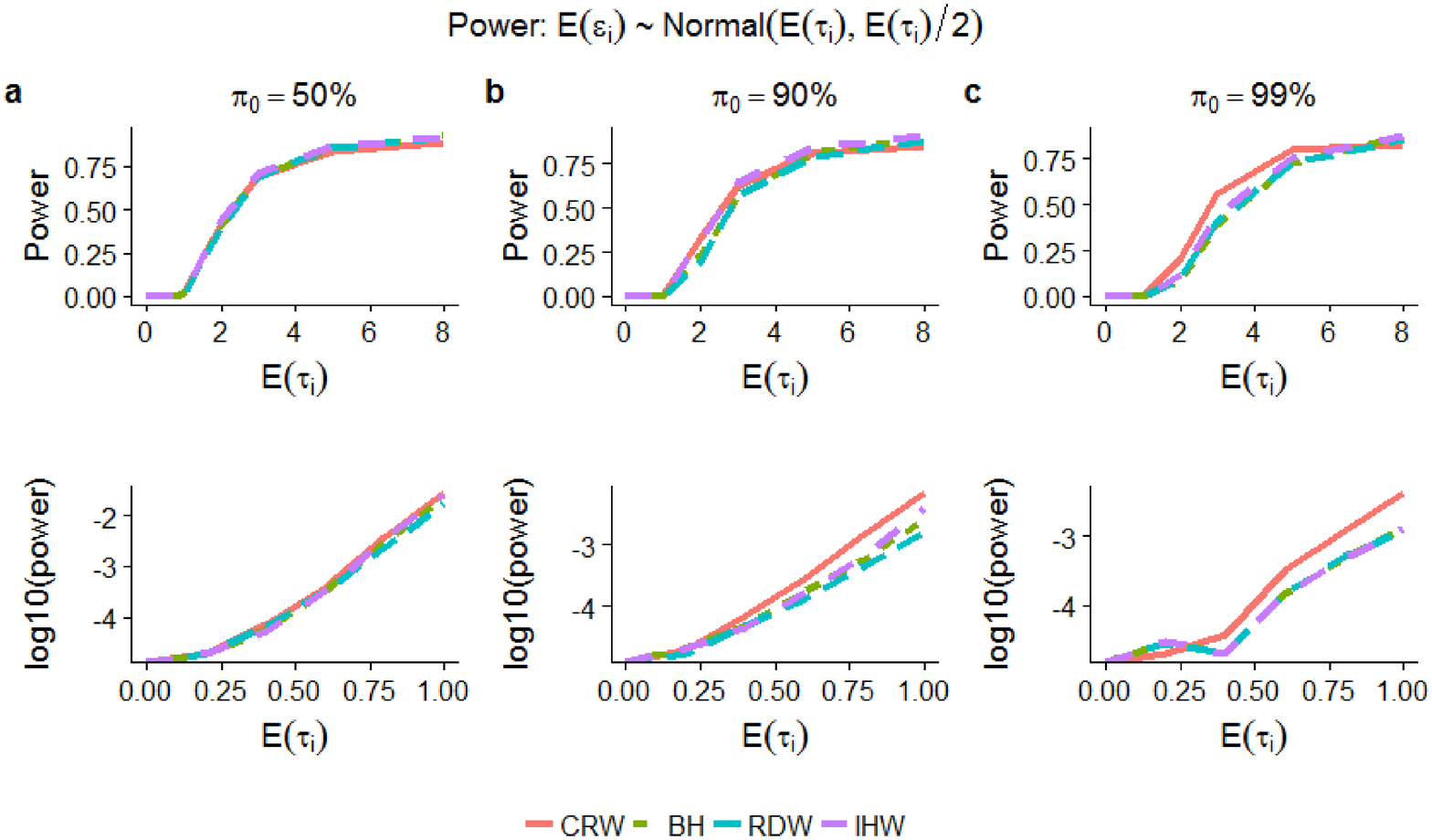}
	% note that files may not be rotated
	\caption{The power for the same parameters described in Fig.~\ref{fig:power1} except the mean test-effect 
	$E(\varepsilon_i)$ is not equal to the mean covariate-effect $E(\tau_i)$; rather $E(\varepsilon_i) \sim 
	Normal\big(E(\tau_i), E(\tau_i)/2\big)$, where $CV=1/2$ (for the other cases see SI).}
	\label{fig:power2}
\end{figure}

In the above simulations, we assume that there is a noiseless, linear relationship between the test effect 
size and the covariate effect size. Next, we show simulations that explore the effect of noise in the test 
effect-covariate effect relationship. We assume that the test effect is normally distributed around the 
covariate effect. In the simulation shown (Figure~\ref{fig:power2}), the standard deviation is chosen so that 
the coefficient of variation (CV) is $1/2$. That is, the standard deviation in test effect size about the 
covariate effect size is $1/2$ the value of the true covariate effect size. This scenario is typical of high 
throughput data. There is not a large effect on power for any of the methods and CRW is still the most 
effective. Further simulations are shown in the SI. All of the methods lose effectiveness as the noise 
increases, but the relative benefit of CRW decreases with larger noise and CRW eventually becomes inferior. 
However, this does not occur until the level of noise in the test effect – covariate effect relationship is 
such that no covariate-based method will be effective.

In calculating CRW weights, we assume independence between hypothesis tests. Simulations (see SI) show that 
correlation between tests does not have a large impact on power, either in absolute terms or comparisons 
between methods. The exception is for very high correlation $(>80\%)$, at which the performance of IHW 
increases relative to CRW.

\section{Data application}
\begin{figure}
	% The arguments in the next line are {height}{optional width}{used only by OUP for typesetting}[filename, 
	%in directory art]
	%\figurebox{20pc}{25pc}{}[realdataexamples.eps]
	\includegraphics[width=5.5in,height=2.75in]{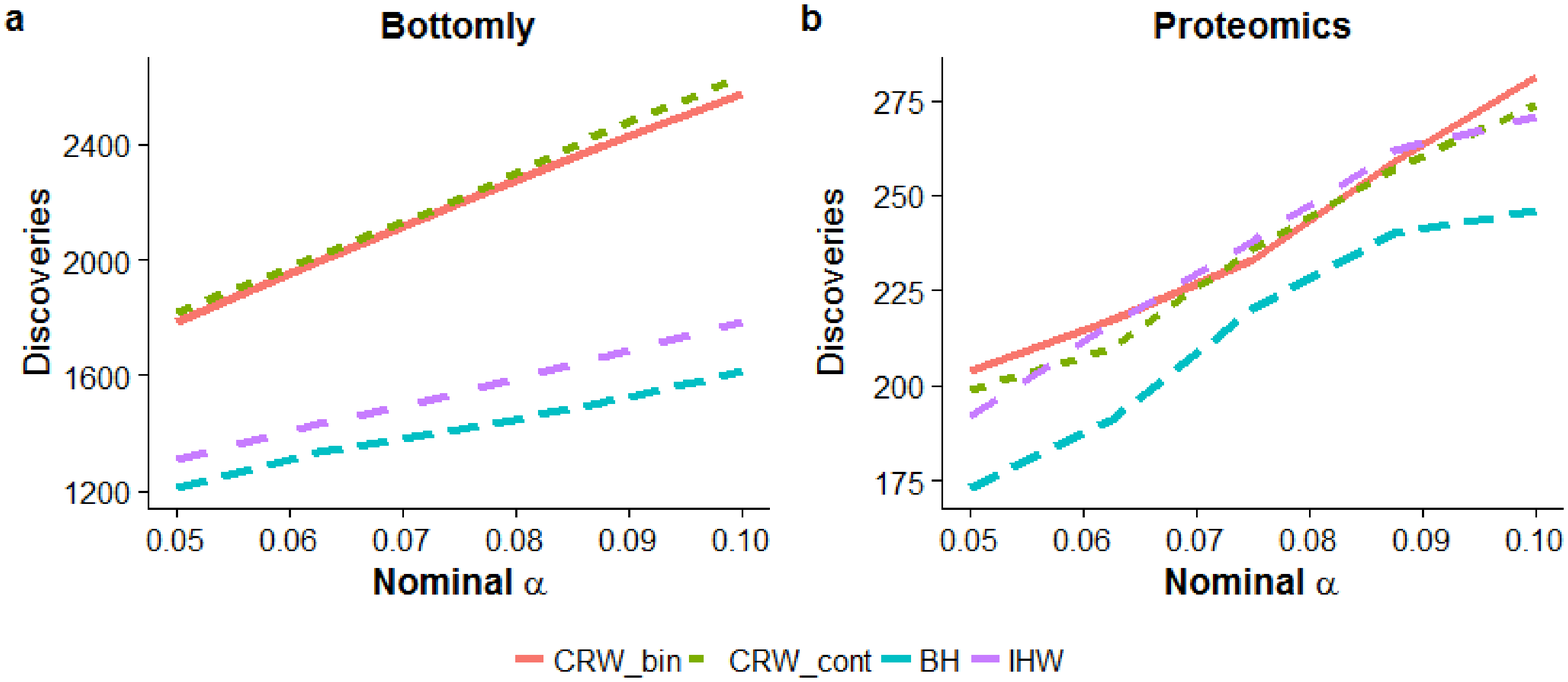}
	% note that files may not be rotated
	\caption{The number of rejected null hypotheses across different significance levels of $\alpha$. Here, 
	CRW\_bin =  CRW for binary effects, CRW\_cont =  CRW for continuous effects, BH = Benjamini-Hochberg, and 
	IHW = Independent Hypothesis Weighting methods.}
	\label{fig:real_data_examples}
\end{figure}

We show two data examples (Figure~\ref{fig:real_data_examples}): 1) Bottomly--an RNA-Seq dataset 
\citep{bottomly2011evaluating}, which has $16,183$ genes and was downloaded from the Recount project 
\citet{frazee2011recount} and 2) Proteomics dataset \citet{dephoure2012hyperplexing} in which differential 
abundance of $2,666$ proteins was tested.

For the Bottomly data, the estimated proportion of the true null hypothesis was $82\%$. The mean test effect 
size for true effects was estimated at $1.7$ and $2.2$ for the binary and continuous cases, respectively. The 
mean covariate effect size for true effect was estimated as $1.2$ for both the binary and continuous cases. 
The IHW method finds approximately $10\%$ more discoveries than BH, whereas CRW finds greater than $50\%$ 
more discoveries. Fig.~\ref{fig:hightroughputPower}a shows that there are many effects with very low power. 
The ability of CRW to boost the power for weak effects results in a large gain in discoveries.
For the Proteomics data, the estimated proportion of the true null hypothesis was $66\%$. The mean test 
effect size for true effects was estimated as $1.8$ and $2.0$ for the binary and continuous cases, 
respectively. The mean covariate effect size for true effect was estimated as $0.7$ for both the binary and 
continuous cases. In this case, IHW and CRW have similar performance, finding about $10\%$ more discoveries 
than BH. This dataset has conditions that are more favorable for the IHW method relative to CRW: the 
estimated proportion of true nulls is rather low $(66\%)$ and the power at average effect size is higher than 
for the Bottomly data.

\section{Discussion}
The multiple testing problem is fundamental to high throughput data, and the necessary statistical stringency 
makes detection of features of interest difficult unless their effect size is large. We have shown here for 
representative data sets that the power to detect most true effects is very low. Unless this problem can be 
overcome, then the promise of high throughput data may never be realized. Weighted p-values provide a 
framework for using external information to prioritize the data features that are most likely to be true 
effects. Many studies 
\citep{spjotvoll1972optimality,holm1979simple,benjamini1997false,satagopan2002two,rubin2006method,roeder2006using,wasserman2006weighted,ionita2007genomewide,farcomeni2008review,roeder2009genome,roquain2009optimal,bourgon2010independent,dobriban2015optimal,kim2015prioritizing}
 have proposed methods of p-value weighting and techniques have steadily advanced. However, all existing 
methods require either 1) difficult to attain knowledge of effect sizes or effect size distributions or 2) 
grouping tests by the covariate and then estimating properties of the groups. While group-based methods are 
powerful in many scenarios, their effectiveness may suffer when true features of interest are rare and/or 
have effect sizes such that power will be low.

Very low statistical power for detection of most features of interest is the norm for high throughput 
biological data. Furthermore, most tests in most datasets are null. This was observed for both the RNA-seq 
and proteomics datasets used in this paper, and our experience is that this is typical. The CRW method 
proposed here assigns an individual weight for each test based on its ranking by the external covariate and 
gives the strongest weights to true effects when they are weak and rare. It is maximized at the most probable 
rank for a test that has the average effect size among true effects. Thus, the highest weight tends to be at 
the center of the distribution of effect sizes for true effects. These weights will not be washed out by 
group effects. In contrast, in group-based methods like IHW, all covariate-ordered groups will have many null 
tests, and thus weights will tend to be diluted. We have shown that, as a result, discovery probabilities for 
weak effects can be increased by as much as $10$-fold relative to the state-of-the-art method IHW.

Although true effects typically form a small proportion of all tests, there are still a large number of them 
in high throughput datasets. Even though they individually have small power, we expect to detect some because 
of their large numbers. Thus, the large increase in power relative to other methods that CRW exhibits at low 
effect sizes can translate into large numbers of extra discoveries even though absolute power is still low.

The CRW method requires knowledge of the probabilistic relationship between covariate rank and test effect 
size. In this paper, we assume that the covariate follows a normal distribution and then derive an expression 
for the distribution of covariate rank conditioned on covariate effect size. Performance may suffer when the 
covariate deviates substantially from normality. The covariate rank-test effect relationship can also be 
estimated empirically, although the benefit relative to IHW decreases \citep{hasan2017optimal}.

Like all covariate-based methods, CRW depends crucially on how well the covariate identifies features of 
interest. We assume that the test effect size and covariate effect size are related by a linear model. As the 
relationship gets noisier, the effectiveness of the weights decreases. Simulation results show that power is 
not highly affected if the coefficient of variation of the test effect around the covariate effects is less 
than about one. The effectiveness decreases as the coefficient of variation increase beyond one. A 
fundamental question for all covariate-based methods is whether covariates exist that are able to prioritize 
promising features sufficiently well to be useful. An important task for future research is to determine good 
covariates for different data types and how effective those covariates are. Better knowledge of covariates is 
essential to apply weighting methods effectively and is an important avenue of research for optimization of 
power in high throughput studies.

\appendix
%\appendixone
\section{Appendix}
\subsection{Proof of theorem \ref{CovRank_thm} - covariate rank}
\begin{proof}
	Consider a multiple hypothesis testing situation where $m$ tests are being conducted. Each test $i$ has 
	an associated covariate value with effect size $\tau_i$. Suppose there are $m_0$ null covariate values 
	with $\tau_i=0$ and $m_1=m-m_0$ true alternative covariate values with $\tau_i>0$. Let, 
	$x_1,\ldots,x_{m_0}$ be the covariate values for which the null hypothesis is true and $y_1,\ldots,y_{m_1 
	}$ be the covariate values for which the alternative hypothesis is true, and also consider a covariate 
	value $t$. Our goal is to find the distribution for the rank of the covariate value $t$ among the other 
	covariates.
	Let $k_0-1$ and $k_1-1$ refer to the number of success, i.e., the number of covariates greater than $t$ 
	among the null and alternative covariates, respectively.
	Define, $I_j$ and $I_l$ as follows
	\begin{equation*}
		I_j=
		\begin{cases}
			1, & \text{if } X_j > t\\
			0, & \text{otherwise}
		\end{cases}; \ j = 1,\ldots,m_0
	\end{equation*}
	and		
	\begin{equation*}
		I_l=
		\begin{cases}
			1, & \text{if } Y_l > t\\
			0, & \text{otherwise}
		\end{cases}; \ l = 1,\ldots,m_1.
	\end{equation*}	
	Take $S_0 = \sum_{j=1}^{m_0} I_j$, then $S_0 + 1$  will give the rank of $t$ among the null covariates. 
	$I_j$ is a binary random variable, and assuming independence among the covariates, the number of 
	covariates exceeding $t$ follows a binomial distribution with $m_0$ trials and success probability 
	$P(X_j>t)$. $P(X_j>t)$ can be computed as $P(X_j>t)=1-P(X_j<t)=1-F_0 (t)$, where $F_0 (.)$ is the 
	cumulative density function of $X_j$  under the null model. For simplicity, we will denote it $F_0$. 
	Thus, if $k_0$  is the rank of $t$ denoted by a random variable $r_{0j}$, then
	\begin{equation*}
		P(r_{0j}=k_0 \mid t )=\binom{m_0}{k_0-1} (1-F_0 )^{k_0-1} F_0^{m_0-(k_0-1)}; \ 1 \le k_0 \le m_0.
	\end{equation*}
	Similarly, for the alternative hypotheses we have $S_1 = \sum_{l=1}^{m_1} I_l$; therefore, the 
	probability of success is $P(Y_l>t)=1-P(Y_l<t)=1-F_1 (t)=1-\int F_1 (t,\tau_l)f(\tau_l)d\tau_l$. $F_1(t)$ 
	is the probability of a randomly chosen covariate statistic exceeding $t$. This depends on the effect 
	size $\tau_l$. This effect size is unknown and thus we integrate over possible values. $F_1 (t,\tau_l)$ 
	refers to the cumulative density function of $Y_l$, with $\tau_l$ known.	We will denote $P(Y_l<t)$ as 
	$F_1$. Thus, if $k_1$  is the rank of $t$ denoted by a random variable $r_{1l}$, then
	\begin{equation*}\label{eq:contAltRanksProb}
		P(r_{1l}=k_1 \mid t)=\binom{m_1}{k_1-1} (1-F_1 )^{k_1-1} F_1^{m_1-(k_1-1)}; \ 1 \le k_1 \le m_1.
	\end{equation*}
	In practice, we do not know $m_0$ and $m_1$. These parameters can be estimated by the method of 
	\cite{storey2003statistical}. 
	
	Our goal is to obtain rank $k$ under both the null and alternative models, simultaneously, for a random 
	covariate $t$. If the rank of $t$ is $k$, then there are $k-1$ covariates that are higher and $m-k$ 
	covariates that are lower than the covariate $t$, assuming that $t$ is counted as one of the $m$ 
	covariates. Let us suppose a random variable $r_i$ refers to the rank of $t$ among all (null and 
	alternative) covariate values simultaneously. Then
	\begin{equation*}
		P(r_i=k \mid \tau_i )=P(r_{0j}+ r_{1l}-1=k \mid \tau_i ),
	\end{equation*}
	where $i=1,\ldots,m; \ 0 \le k \le m$; and $m_0+m_1=m$. This is a joint distribution of $r_{0j}$ and 
	$r_{1l}$. Thus, by applying convolution for the discrete case, we have
	\begin{equation}\label{eq:ranksProbFirst}
		P(r_i=k \mid \tau_i )=\frac{\sum_{k_0=1}^{k}P(r_{1l}+r_{0j}=k+1,\tau_i,r_{0j}=k_0 ) }{p(\tau_i)}.
	\end{equation}
	Up to this point $t$ has been assumed to be a fixed value. Now, we let $t$ vary with distribution $P(t 
	\mid \tau_i)$.  Incorporating this into (\ref{eq:ranksProbFirst}) produces
	\begin{equation}\label{eq:ranksProbFinal}
		P(r_i=k \mid \tau_i)=\sum_{k_0=1}^{k} \int_t P(r_{1i}=k-k_0+1 \mid \tau_i,t) P(r_{0i}=k_0 \mid 
		\tau_i,t )P(t \mid \tau_i)dt.
	\end{equation}
	If $t$ is one of the $m_0$ null covariates, and we want to compute $P(r_{0i}=k_0 \mid \tau_i,t)$, then 
	the number of null trials in the binomial is $m_0-1$ and the number of alternative trials is $m_1$. If 
	$t$ is one of the $m_1$  alternative covariates, then the number of null trials is $m_0$  and the number 
	of alternative trials is $m_1-1$. Similar arguments hold for calculating $P(r_{1i}=k-k_0+1 \mid 
	\tau_i,t)$. Therefore,
	\begin{equation*}
		P(r_{1i}=k-k_0+1 \mid \tau_i,t)=
		\begin{cases}
			\binom{m_1}{k-k_0} (1-F_1 )^{k-k_0} F_1^{m_1-(k-k_0)}, & \text{if } \tau_i = 0\\
			\binom{m_1-1}{k-k_0} (1-F_1 )^{k-k_0} F_1^{(m_1-1)-(k-k_0)}, & \text{if } \tau_i > 0
		\end{cases},
	\end{equation*}
	and
	\begin{equation*}
		P(r_{0i}=k_0 \mid \tau_i,t)=
		\begin{cases}
			\binom{m_0-1}{k_0-1} (1-F_0 )^{k_0-1} F_0^{m_0-k_0}, & \text{if } \tau_i = 0\\
			\binom{m_0}{k_0-1} (1-F_0 )^{k_0-1} F_0^{m_0-(k_0-1)}, & \text{if } \tau_i > 0
		\end{cases}.
	\end{equation*}
	The expression $P(r_i=k \mid \tau_i)$ is, in fact, the required probability of rank of the covariate 
	given the effect size. Since $P(t\mid \tau_i )$ is the probability density function of $t$, 
	(\ref{eq:ranksProbFinal}) can be simplified in terms of expectation ($E_T$) over the random variable $t$ 
	as
	\begin{equation*}
		P(r_i=k \mid \tau_i )=\sum_{k_0=1}^{k}E_T \Big[ P(r_{1i}=k-k_0+1 \mid \tau_i,t )P(r_{0i}=k_0 \mid 
		\tau_i,t )\Big]
	\end{equation*}
	This finishes the proof.
\end{proof}

\subsection{Proof of proposition \ref{CovRankProp} - covariate rank is the expected value of the normal PDF}
\begin{proof}
	If $\tau_i = 0,$ then from (\ref{eq:ranksProbFinal}) $P(r_i=k \mid \tau_i)$ can be written as
	\begin{equation*}
		P(r_i=k \mid \tau_i) = \sum_{k_0=1}^{k} E_T 
		\bigg[\binom{m_0-1}{k_0-1}(1-F_{0})^{k_0-1}F_{0}^{m_0-k_0}.\binom{m_1}{k-k_0}(1-F_{1})^{k-k_0}F_{1}^{m_1-(k-k_0)}
		 \bigg].
	\end{equation*}
	Equivalently,
	\begin{equation*}
		P(r_i=k \mid \tau_i) = E_T \sum_{k_0=1}^{k} 
		\bigg[\binom{m_0-1}{k_0-1}(1-F_{0})^{k_0-1}F_{0}^{m_0-k_0}.\binom{m_1}{k-k_0}(1-F_{1})^{k-k_0}F_{1}^{m_1-(k-k_0)}
		 \bigg].
	\end{equation*}
	Suppose we have two Binomial random variables $X=k_1-1\sim Binom(m_1,1-F_1)$ and $Y=k_0-1 \sim 
	Binom(m_0-1,1-F_0)$; and we want to obtain \textit{pmf} of $Z=X+Y$. Then
	\begin{equation*}
		\begin{split}
			P(Z=k-1) &=P(X+Y=k-1)\\
			&=P(X=k-1-Y)\\
			&=\sum_{k_0-1=0}^{k-1}P(X=k-k_0)P(Y=k_0-1)\\
			&=\sum_{k_0-1=0}^{k-1}\binom{m_1}{k-k_0}(1-F_{1})^{k-k_0}F_{1}^{m_1-(k-k_0)}.\binom{m_0-1}{k_0-1}(1-F_{0})^{k_0-1}F_{0}^{m_0-k_0}.
		\end{split}
	\end{equation*}
	Therefore,
	\begin{equation*}
		P(r_i=k \mid \tau_i)=E_T P(Z=k-1)=E_T P(Z+1=k),
	\end{equation*}
	where $Z$ is a sum of two independent binomial random variables. Thus, the PDF of $Z+1$ can be 
	approximated by normal PDF with mean $= E(X+Y+1)$ and variance $= var(X+Y+1)$. Similarly, we can also 
	obtain an approximation when $\tau_i > 0$.
\end{proof}

\bibliographystyle{Chicago}

\bibliography{References}
\end{document}